\documentclass[twocolumn,aps,amsmath,amssymb,showpacs]{revtex4}

\usepackage{graphicx} 
\usepackage{epsfig}

\begin{document}

\title{On the diffusive anomalies in a long-range Hamiltonian system}

\author{Luis G. Moyano}
\affiliation{Centro Brasileiro de Pesquisas F\'{\i}sicas - Rua Xavier Sigaud
  150, 22290-180, Rio de Janeiro, Brazil
}

\author{Celia Anteneodo}
\affiliation{Departamento de F\'{\i}sica, 
Pontif\'{\i}cia Universidade Cat\'olica do Rio de Janeiro,  
CP 38071, 22452-970, Rio de Janeiro, Brazil}


\begin{abstract}
We scrutinize the anomalies in diffusion observed in an  
extended long-range system of classical rotors, the HMF model. 
Under suitable preparation, the system falls into long-lived quasi-stationary 
states presenting super-diffusion of rotor phases. 
We investigate the diffusive motion of phases by monitoring the 
evolution of their probability density function for large system sizes. 
These densities are shown to be of the $q$-Gaussian form, 
$P(x)\propto (1+(q-1)[x/\beta]^2)^{1/(1-q)}$, with parameter $q$ 
increasing with time before reaching a steady value $q\simeq 3/2$. 
From this perspective, we also discuss the relaxation to equilibrium and show 
that diffusive motion in quasi-stationary trajectories strongly 
depends on system size. 
\end{abstract}

\maketitle

\section{Introduction}

Systems with long-range interactions constitute a very appealing subject 
of research as they display a variety of dynamic and thermodynamic 
features very different from those of short-range systems treated in 
the text-books 
(see \cite{lrange} for a review on the subject). 
Moreover, in recent years, the study of long-range models have raised a 
renewed interest due to the possible applicability of 
``Nonextensive Statistics''\cite{tsallis} to such systems.

A very simple model that offers the possibility of investigating many issues 
related to long-range interactions is the Hamiltonian Mean-Field 
(HMF) model\cite{hmf}.
It consists of $N$ planar classical spins interacting through infinite-range 
couplings. The dynamical variables of each spin $i$ are a phase angle
$\theta_i$ and its conjugate momentum $p_i$ whose evolution derives from 
the Hamiltonian 
\begin{equation} \label{HMF}
H = \frac{1}{2}\sum_{i=1} ^{N} p_{i} +
\frac{1}{2N} \sum_{i,j=1} ^{N}
\left[1-\cos(\theta_{i}-\theta_{j})\right].
\end{equation}
This model can be seen as a variant of the $XY$ ferromagnet, 
equipped with a natural Newtonian dynamics. 
Although the range of  interactions is infinite, the HMF has been 
shown to behave in many aspects 
qualitatively like its long(finite)-range analogs\cite{alfaxy}.
Then, despite its simplicity, it reflects features of real systems 
with long-range forces such as galaxies and charged plasmas\cite{lrange}. 

Its equilibrium thermodynamics can be solved in the canonical ensemble. 
It presents a ferromagnetic second-order phase transition, from a low energy 
clustered phase to a high energy homogeneous one that occurs at critical 
temperature $T_c=0.5$ (i.e., critical specific energy 
$\varepsilon_c=0.75$)\cite{hmf}. 
However, concerning the microscopic dynamics,  
the system may get trapped into 
trajectories over which averaged quantities remain constant 
during long periods of time 
with values different from those expected at equilibrium.  
For instance, for {\em water-bag} initial conditions 
(i.e., $\theta_i=0$ $\forall i$, 
and $p_i$ randomly taken from a uniform distribution), a quasi-stationary 
(QS) state appears at energies close below $\varepsilon_c$\cite{qss}. In a QS state, 
the temperature (twice the specific mean kinetic energy) is almost constant in time 
and lower than the canonical value to which it eventually relaxes. However, 
the duration of  QS states increases with the system size $N$, indicating that 
these states are indeed relevant in the ($N\rightarrow\infty$) thermodynamical 
limit (TL).

Several other peculiar features have been found for   
out-of-equilibrium initial conditions, 
e.g., negative specific heat\cite{hmf}, 
non-Maxwellian momentum distributions\cite{qss}, glassy dynamics\cite{glassy},
aging\cite{aging1,aging2}, 
anomalous diffusion\cite{superdif}, and others. 
In particular, anomalous diffusion has been initially associated to 
quasi-stationarity\cite{superdif} and later to the (non-stationary) relaxation to 
equilibrium\cite{yamaguchi}. Moreover, controversies about the 
characterization of these states have arisen\cite{montemurro},  
remaining still aspects of the anomaly in diffusion to be 
investigated. 
In this work we report new results on anomalous diffusion 
and relaxation to equilibrium, focusing on the dependence on system size.

\section{Results}

The equations of motion derived from Eq.~(\ref{HMF}) 
\begin{eqnarray} \nonumber
\dot{\theta_i} &=& p_i \,, \;\;\;\;\;\;\mbox{for $1\le i\le N$}\\ 
\dot{p_i} &=&  M_y \cos{\theta_i} - M_x\sin{\theta_i},  \label{newton}
\end{eqnarray}
where $\vec{M} =\frac{1}{N}\sum_j (\cos{\theta_j},\sin{\theta_j})$ 
is the magnetization, were solved by means of a symplectic fourth order 
algorithm \cite{yoshida}. 
Integrations were performed for fixed $\varepsilon=0.69$, for which 
quasi-stationary effects are more pronounced. 
Moreover, in the continuum limit, QS trajectories at energies around that 
value are stable stationary solutions 
of Vlasov equation\cite{vlasov,vlasov2}. 
We considered two classes of initial conditions. 
One of them is a slight
variation of water-bag initial conditions normally used in the literature: 
setting $\theta_i=0$ for $i=1,..., N$, and regularly valued momenta (instead of 
random ones)  
with the addition of a small noise to allow for statistical
realizations. This type of initial condition preserves the main features 
of random water-bag ones but the QS temperature $T_{QS}$ is even 
lower and the duration
of the QS  regime is longer, in some way mimicking larger systems\cite{vlasov}. 
We also performed simulations for equilibrium (EQ) initial conditions: waiting 
an appropriate transient after picking spatial coordinates and momenta from 
Boltzmann-Gibbs statistics.

Integration of Eqs.~(\ref{newton}) yields phases in $(-\infty,\infty)$. 
Then, we measured the histograms of phases at different times, 
accumulating the data over several realizations to 
improve the statistics. 
Actually, the dynamics depends on the phases modulo $2\pi$, since they 
enter into the equations of motion as 
argument of sine and cosine  functions. 
However, the statistics of unbounded phases is relevant as it 
reflects features of momentum space, which displays anomalies such as 
non-Maxwellian distributions\cite{qss} and two-time correlations signaling 
``aging'' effects\cite{aging2}. 
For instance, a simple mathematical relationship exists between the standard 
deviation of angles and velocity correlations\cite{yamaguchi}.

\begin{figure}[bh!] 
\begin{center} 
\includegraphics*[bb=84 445 514 735, width=0.5\textwidth]{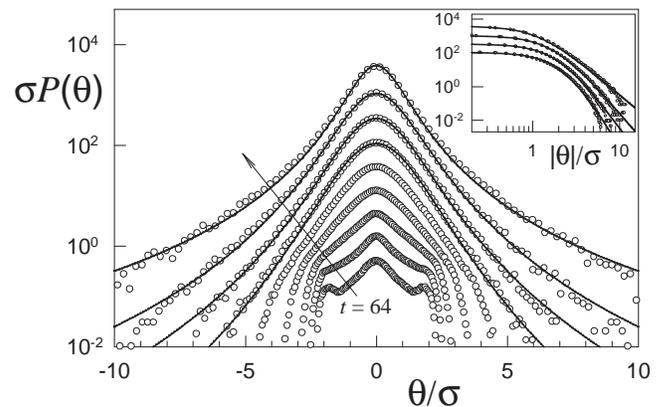} 
\caption{Histograms of rotor phases at different instants of the dynamics 
(symbols). 
Simulations for $N=1000$ were performed starting from 
regular water-bag initial conditions at $\varepsilon=0.69$ 
(conditions leading to QS states). 
Countings were accumulated over $100$ realizations, at 
times $t_k=2^k$, with $k=6,8,..14$, growing in the direction 
of the arrow up to $t=16384$. 
Solid lines correspond to $q$-Gaussian fittings. 
Histograms were shifted for visualization. 
Inset: log-log representation of the fitted data. 
}
\label{fig:pdfs_wb}
\end{center}
\end{figure}

For low energies, phases are confined, while for sufficiently large energies, 
they evolve in diffusive motion. 
In Fig.~\ref{fig:pdfs_wb} we show the properly scaled probability density functions 
(PDFs) of rotor phases at different times, for fixed size $N=1000$ and 
energy per particle $\varepsilon=0.69$, starting from water-bag initial conditions. 
After transient stages, numerical histograms 
can be very well described in the whole range 
by the $q$-Gaussian function \cite{tsallis}  
\begin{equation} \label{pdf}
P(\theta)=A\left( 1+ (q-1) (\theta/\beta)^2 \right)^\frac{1}{1-q} \, ,
\end{equation}
where $A$ is a normalization factor and $\beta$ a positive constant. This
function includes the standard Gaussian when
$q\rightarrow\,1$ and presents power-law tails for $q>1$. 
Recalling that the PDF~(\ref{pdf}) has variance 
$\sigma^2=\beta^2/(5-3q)$, for $q<5/3$, 
and considering normalized angles $\phi=\theta/\sigma$, then Eq.~(\ref{pdf}) 
can be written as the uni-parametric function 

\begin{equation}
P_q(\phi)=A_q \left( 1+\frac{q-1}{5-3q}\,{\phi}^2 \right)^\frac{1}{1-q},  
\end{equation}
with $A_q=\sqrt{\frac{q-1}{\pi(5-3q)}}
\frac{\Gamma(1/(q-1))}{\Gamma(1/(q-1)-1/2)}$. 
At each instant $t$ of the dynamics, $\sigma^2$ is computed as 
\begin{equation} \label{variance}
\sigma^2(t) = \langle(\theta-\langle\theta\rangle_t)^2\rangle_t,
\end{equation}
where $\langle...\rangle_t$ denotes average over the $N$ rotors and over 
different realizations of the dynamics at time $t$. 
As far as we know, this is the first time that pure 
$q$-Gaussian PDFs arising from a Hamiltonian dynamics are detected.

\begin{figure}[h!] 
\begin{center}
\includegraphics*[bb=104 217 540 750, width=0.5\textwidth]{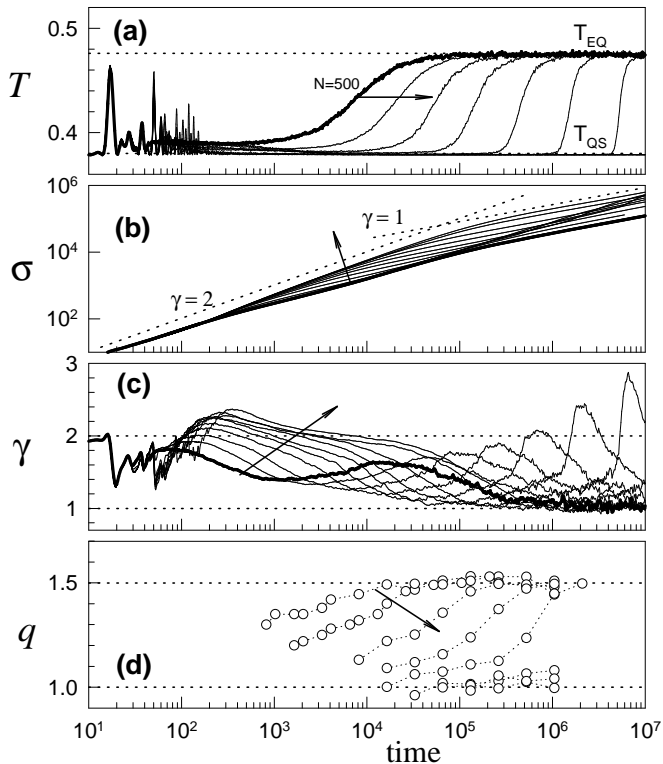} 
\caption{Averaged time series of (a) temperature $T$, 
 (b) deviation $\sigma$,  (c) diffusion exponent $\gamma$   
and  (d) parameter $q$, for $\varepsilon=0.69$ and different values of $N$ 
($N=500\times 2^k$, with $k=0,\ldots,9$).  
Bold lines correspond to $N=500$, as reference, and $N$ increases in the 
direction of the arrows up to $N=256000$.
Averages were taken over $2.56\times 10^5/N$ realizations, 
starting from a {\em waterbag} configuration at $t=0$. 
In panel (d), the fitting error is approx. 0.03. 
Dotted lines are drawn as references. In (a), they correspond to 
temperatures at equilibrium ($T_{EQ}=0.476$) and at QS states in the TL 
($T_{QS}=0.38$). In (b), to ballistic motion ($\gamma=2$) and normal 
diffusion ($\gamma=1$). }
\label{fig:qss}
\end{center}
\end{figure}

\begin{figure}[h!] 
\begin{center}
\includegraphics*[bb=104 331 550 745, width=0.5\textwidth]{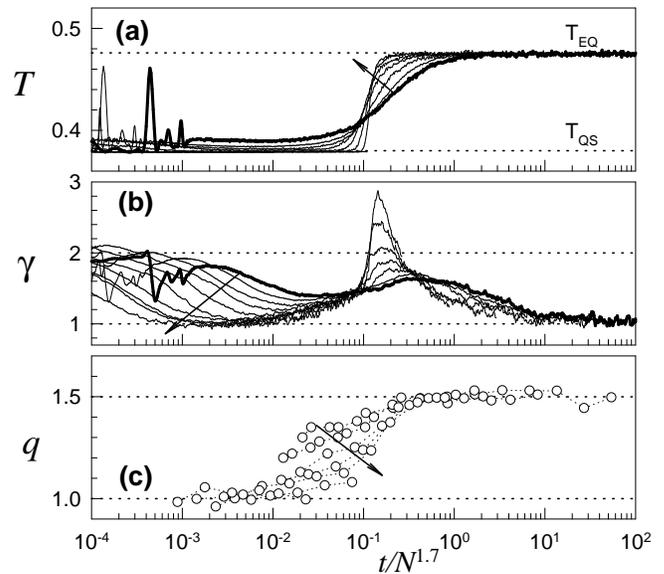} 
\caption{Averaged time series of (a) temperature $T$,   
(b) local exponent $\gamma$  and parameter $q$ (symbols), as a function of 
$t/N^{1.7}$. Data are the same presented in Fig.~\ref{fig:qss}.
}
\label{fig:qss_scaled}
\end{center}
\end{figure}

In Fig.~\ref{fig:qss}d, the evolution of $q$ for different $N$ is 
displayed, including the 
fitting values used in Fig.~\ref{fig:pdfs_wb} for $N=1000$. 
Parameter $q$ increases up to a steady value in the long-time limit, that 
for all $N$ falls within the range $q\simeq 1.51\pm 0.02$. 
In the same figure, we also 
exhibit the temporal evolution of temperature  
$T(t)=\sum_i \langle p^2_i \rangle/N $ (Fig. \ref{fig:qss}.a) 
and dispersion $\sigma(t)$ (Fig. \ref{fig:qss}.b), in order to 
establish a parallel between the different stages of the relevant quantities. 
Instead of single runs, averaged quantities are presented.  
Since they are much less noisy, they allow the 
representation of several curves in the same plot without 
distorting the mean features observed in individual runs.   
We can see that $q$ attains a steady value 
approximately when the QS$\to$EQ transition is completed.
In Fig.~\ref{fig:qss_scaled} the same data are presented for rescaled times. 

\begin{figure}[h!] 
\begin{center} 
\includegraphics*[bb=70 435 514 720, width=0.5\textwidth]{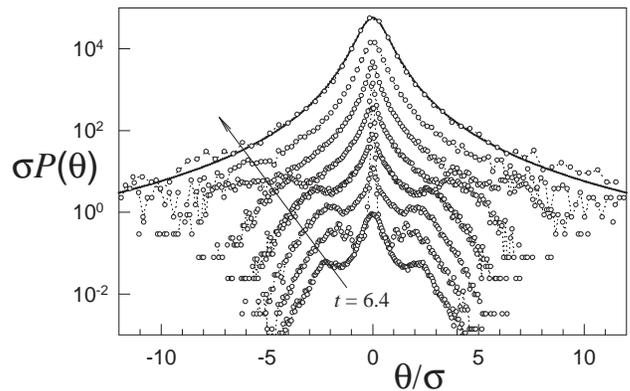} 
\caption{Histograms of rotor phases at different instants of 
the dynamics (symbols). 
Simulations were performed for $N=500$ and $\varepsilon=0.69$, 
starting from an equilibrated initial condition. 
Countings were accumulated over 200 realizations, at 
times  $t_k=0.1\times4^k$, $k=3,4,..10$, 
growing in the direction of the arrow up to $t\simeq 1.05\times10^5$). 
The $q$-Gaussian function with $q=1.53$ was plotted for comparison 
(solid line). 
Histograms were shifted for visualization. 
}
\label{fig:pdfs_eq}
\end{center}
\end{figure}

The PDFs of rotor phases, for $N=500$, and $\varepsilon=0.69$, but 
starting from an equilibrium configuration are shown in Fig.~\ref{fig:pdfs_eq}. 
In this case, histograms present pronounced shoulders that 
persist for long times and 
can not be well described by $q$-Gaussian functions. 
However, as times goes by, the shoulders shift away from the center 
and the histograms tend to a $q$-Gaussian function, 
with $q$ going to $q\simeq 3/2$ from above in the long-time limit. 

Then, starting both from equilibrium and out-of equilibrium (water-bag)
configurations, initially confined densities of phases develop power-law tails 
and adopt $q$-Gaussian forms.

\begin{figure}[h!] 
\begin{center}
\includegraphics*[bb=80 300 530 770, width=0.5\textwidth]{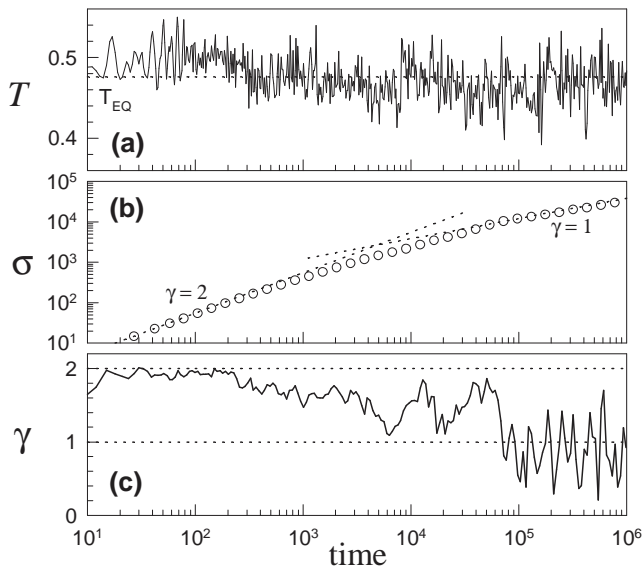} 
\caption{Time series of  (a) temperature $T$,  (b) deviation $\sigma$ and 
(c)  diffusion exponent $\gamma$, for $N=500$ and $\varepsilon=0.69$.  
At $t=0$, the system is in an {\em equilibrium} configuration. 
The outcome of a single run is exhibited.}
\label{fig:eq}
\end{center}
\end{figure}

Diffusion of spatial coordinates may be characterized by the average squared 
displacement $\sigma^2(t)$ of the angles $\theta$ as defined by 
Eq. (\ref{variance}). 
In the one-dimensional generalized Einstein relation  
\begin{equation}
\sigma^2(t)  =  2D\, t^\gamma\, ,
\label{diffusion}
\end{equation}
where $D$ is the diffusion constant, 
the case $\gamma=1$ corresponds to normal diffusion, $\gamma<1$ to
sub-diffusion and super-diffusion occurs for $\gamma>1$. 
The evolution of $\sigma$ is shown in Figs.~\ref{fig:qss}b and \ref{fig:eq}b, 
for water-bag and equilibrium initial preparations, respectively. 
In order to detect different regimes, it is useful to obtain 
an instantaneous exponent $\gamma$ as a function of
time by taking the logarithm in both sides of Eq.~(\ref{diffusion}) and
differentiating with respect to $\ln t$:
\begin{equation}
\gamma(t) = \frac{d(\ln \sigma^2)}{d(\ln t)}\,\,.
\label{gamma}
\end{equation}
The outcome of this procedure can be seen in Fig.~\ref{fig:qss}c for water 
bag initial conditions.
The same analysis for systems prepared in an equilibrated configuration is 
presented in Fig.~\ref{fig:eq}c.

For the latter class of initial conditions, phase motion is 
ballistic ($\gamma=2$) at short time scales in which rotors move almost 
freely, while phases display normal diffusion ($\gamma=1$) 
in the long-time limit, 
as shown in Fig.~\ref{fig:eq} (see also \cite{yamaguchi}).  
The crossover between both behaviors shifts to larger times as $N$ increases. 
We observed this same pattern of behaviors also at supercritical energies 
($\varepsilon=5$), although in Ref. \cite{superdif} only 
the ballistic regime was reported since normal diffusion settles at times 
longer than those analyzed in that work.

For water-bag initial conditions, 
both ballistic and normal regimes are also observed for short and long times, 
respectively. 
However, in this case, the crossover is more complex 
(see Figs.~\ref{fig:qss}(c) and \ref{fig:qss_scaled}(c)):  
Exponent $\gamma$ varies non-monotonously taking super-diffusive 
values and it does not present a well defined plateau as temperature does.  
For this intermediate stage, super-diffusion has been reported before and 
attributed to a kind of L\'evy-walk mechanism, yielding a  
succession of free walks and trapping events\cite{superdif}.  
Also, it has been interpreted from a topological perspective\cite{topo}. 

The main features one observes in the profile of local  exponent 
$\gamma$ vs. time can be 
summarized as follows: 
(i) In a first stage, $\gamma$ takes a maximum value 
that remains close to $\gamma=2$, corresponding to ballistic motion. 
This stage occurs at the beginning of the QS regime and its 
duration increases with $N$. 
(ii) Then $\gamma$ reaches a minimum value that as $N$ increases 
falls within the QS range. 
Also as $N$ increases, the valley broadens and flattens, 
and the minimum value tends to unity (see Fig.~\ref{fig:qss_scaled}).  
Thus, anomalous diffusion in QS states is a finite-size effect, as already 
pointed out by Yamaguchi from a different perspective and 
for a different initial preparation\cite{yamaguchi}. 
(iii) Another maximum appears in correspondence to the rapid relaxation 
from QS to EQ. 
The maximal value grows with $N$, largely exceeding the value $\gamma=2$. 
But the peak narrows, hence it lasts less. 
This peak corresponds approximately to the inflexion point in 
the time evolution of temperature (Figs. \ref{fig:qss} and \ref{fig:qss_scaled}), 
whose slope increases with $N$. 
In fact, in the QS$\to$EQ relaxation, 
the rapid increase of $T$ (kinetic energy) leads to 
an accelerated increase of the phases in average.  
(iv) In the final stage, $\gamma$ relaxes asymptotically to unity, 
indicating normal diffusion at very long times. 

\section{Discussion and remarks}

From the analysis of $\gamma$, one concludes that   
diffusive motion in QS trajectories strongly depends on the system size.  
In the $\gamma$ vs. $t$ profile, as $N$ increases, the neighborhood 
of the first minimum 
(that becomes contained in the QS range defined by $T$) flattens, 
defining a quasi-stationary value of $\gamma$. 
This steady value goes to one in the TL. All these features are neatly 
observed in Fig.~\ref{fig:qss_scaled}.
Then, diffusion in QS states becomes of the normal type in that limit. 
For the whole time span, non-trivial values of $\gamma$ occur in the TL only 
at non-stationary (nor quasi-stationary) stages,  
consistently with observations by Yamaguchi\cite{yamaguchi}. 

Moreover, the spreading of initially confined phases develops power-law tails. 
For QS trajectories, histograms adopt a $q$-Gaussian form with $q$ 
increasing up to a steady value $q\simeq 3/2$. 
A similar steady value is also attained asymptotically when starting 
at equilibrium. 
For both classes of initial conditions,  
stabilization of parameter $q$ is reached only when equilibrium holds  
and $\gamma$ attains the steady value corresponding to normal diffusion. 
 
Let us note that L\'evy densities present power-law tails with exponents 
restricted to a given interval (yielding divergent second moment) 
that in terms of parameter $q$ corresponds to $q>5/3$, therefore, not including 
the observed values. 
While the generalization of the standard diffusion equation with fractional 
spatial derivatives leads to L\'evy functions, the non-linear generalization 
 $\partial_t P(x,t)=D\partial_{xx}[P(x,t)]^{2-q}$, 
with constant $D$ and $q$, has $q$-Gaussian functions as long-time 
solutions\cite{bukman}.  
The non-linear equation yields anomalous diffusion of the correlated type 
with exponent $\gamma=2/(3-q)$.  
In our case, power-law tails also develop. 
Although they are characterized by $q$ changing with time,  
the fact that it attains a steady value suggests that the spreading of phases 
may be ruled by an akin non-linear process. In such case, effective parameters are   
$q_{eff}\simeq 1.5$ and $\gamma_{eff}\simeq 1.33$. 
However, this issue as well as a possible connection with Nonextensive Statistics 
should be further investigated. 

\section*{Acknowledgments}
We are grateful to C. Tsallis for fruitful discussions. 
L.G.M. acknowledges the kind hospitality at Santa Fe Institute where 
part of this work was made. 
We acknowledge Brazilian agency CNPq for partial financial support.

\end{document}